# Influence of ion induced secondary electron emission on the stability of ionisation vacuum gauges

I. Figueiredo[1], N. Bundaleski[1*], O.M.N.D. Teodoro[1]

K. Jousten[2], C. Illgen[2]

[1] *CEFITEC, Department of Physics, Faculty of Sciences and Technology, Nova University of Lisbon, 2829-515 Caparica, Portugal*

[2]*Physikalisch-Technische Bundesanstalt (PTB), Abbestr. 2-12, 10587 Berlin, Germany*

[*]Corresponding author – e-mail address: n.bundaleski@fct.unl.pt

## Abstract

Surface modification of different materials exposed to an environment typical for hot cathode ionisation vacuum gauges was investigated. Such environment has been generated in a specially designed setup which simulates conditions in a Bayard-Alpert ionisation vacuum gauge. Characterisations by means of X-ray photoelectron spectroscopy and work function study have been performed before and after the sample exposure to Ar and $H_2$ gas discharges. The majority of studied materials, such as molybdenum, gold or stainless steel, are considered to be of interest as ion collectors. In addition, experiments with copper and graphite revealed the major processes that are taking place during the exposure to an ionisation vacuum gauge environment. Since the stability of ionisation vacuum gauges can be affected by the change of secondary electron emission properties of ion collectors during the operation, the corresponding electron yield induced by bombardment of low energy $Ar^+$ ions was measured. Results show that exposure to ionisation vacuum gauge environment contributes to development of hydrocarbon layers, independently from the collector material or the dominant species in the gas discharge. Therefore, ion induced secondary electron yield of clean materials will be changing during the gauge operation and eventually reach saturation regardless of the collector material. The results also show that the hydrocarbon layer cannot be desorbed by standard vacuum baking procedures, whilst ion bombardment will even increase the rate of the hydrocarbon layer formation.

# 1. Introduction

The pressure indication in ionisation vacuum gauges (short: ionization gauges) is obtained by measuring a current of ions onto a collector produced from neutral gas molecules by electron impact ionisation. In these devices ions are accelerated towards the collector. Energetic ions impinging on the ion collector surface can cause secondary electron emission, which will increase the measured current according to the following expression: $I_{meas} = I_{ion} \cdot (1 + \gamma_{e,ion})$. The primary ion current is $I_{ion}$, and $\gamma_{e,ion}$ is the ion induced secondary electron yield (IISEY). This yield is surface dependent and one of the major causes of instability of sensitivity of these gauges [1]. For this reason, we investigated the IISEY from materials relevant for ion collector in ionisation vacuum gauges in some detail.

In the energy range of interest (typically up to 250 eV) there are two mechanisms of the secondary electron emission. When an ion approaches few angstrom from a surface, a neutralization accompanied by emission of an Auger electron may take place. The corresponding energetic condition, that the ionisation potential of a projectile is more than two times larger than the work function, is fulfilled by the majority of ion-surface systems. If the electron manages to escape the material, the result is known as potential electron emission since potential energy of an ion was used to excite the electrons [2],[3].

The second mechanism is subthreshold kinetic electron emission. It is based on the electron promotion in a binary collision of impinging ions with target atoms [3],[4],[5]. After such a collision, one of the electrons may end up in a highly excited (autoionising) state, and the deexcitation then takes place via an Auger electron emission. The kinetic energy threshold for such process is defined by the maximum distance of closest approach between the collision partners which will allow an electron promotion.

Due to the high probability of Auger neutralization (inverse of the probability rate is typically about $10^{-15}$ s, being of the same order as the collision time) potential electron emission is apparently restricted to the first atomic layer. The situation is different for subthreshold kinetic electron emission, since the electron promotion may also take place in deeper layers. Ionisation of one of the collision partners is a direct consequence of electron promotion. Ions formed in this way can be neutralized in an Auger process, so that potential electron emission will occur also in the sample interior. Consequently, this type of kinetic electron emission is rather complicated and should be considered as a three-step process (similarly to electron induced secondary electron emission [6]): creation of internal secondary electrons, their transport through the material including energy losses, and electron emission. We stress that the first step is now particularly complex since several phenomena are in play: atomic collision dynamics (including development of cascades of projectile-target and target-target collisions), projectile backscattering into vacuum, electron promotion in a binary collision yielding in eventual electron emission, etc. The situation becomes even more complicated when the sample is not monoatomic. These are probably the reasons why, to our knowledge, a model encompassing the mentioned effects has not yet been developed.

Potential and subthreshold kinetic electron emission can be distinguished by the dependence of $\gamma_{e,ion}$ on the primary projectile energy and charge. Pure potential electron emission does not depend on the kinetic energy. Electron yield caused by kinetic electron emission increases with the kinetic energy. With the energies of ions in an ionisation vacuum gauge, both processes are restricted to the first several atomic layers of a target material. This implies that non-stable surface conditions will be detrimental for $\gamma_{e,ion}$ of ion collector in ionisation gauges. Typical values of IISEY from metals in this energy range span from 1 to 20%, depending on the ion/collector system and on the collector surface condition.

Secondary electron emission from a collector will directly affect gauge sensitivity and cannot be avoided in common ionisation vacuum gauges (e.g. Bayard-Alpert type). Nevertheless, it will not affect gauge stability as long as the collector surface properties are not changing with time. However, the environment in ionisation vacuum gauges is highly reactive, containing not only ions and electrons but also soft X-rays and reactive neutrals. When exposed to such an environment, the collector surface may be changing during the gauge operation, just as its $\gamma_{e,ion}$. The reactive environment may therefore be a serious source of gauge instabilities, i.e. changes of both gauge sensitivities and relative sensitivity factors during the long-term operation. An earlier investigation evidences this change of gauge sensitivity for 10% and 3% in the case of W and Pt-clad Mo ion collectors, respectively, which was attributed to the change of $\gamma_{e,ion}$ [7].

The major goal of this investigation is to get insight into the contribution of $\gamma_{e,ion}$ to the instability of sensitivity of ionisation gauges in order to estimate their measurement uncertainty due to this effect. In addition, we would like to gain information, if a conditioning procedure and/or the selection of collector materials may help to improve the stability of $\gamma_{e,ion}$ in ionisation gauges. To this end, an experimental setup was built enabling exposure of sample surfaces to an environment similar to that, which can be expected in ionisation gauges. This system, that we call ionisation gauge simulator (IGS), was mounted in the preparation chamber of an existing surface science setup, enabling to perform surface characterization by means of X-ray Photoelectron Spectroscopy (XPS), Work Function change (WF) and IISEY measurements before and after the sample exposure to IGS environment without breaking the vacuum envelope. The samples used in this study are technical materials of interest for ion collectors in ionisation gauges (e.g. Mo, Au, stainless steel), prepared in a way typical for production of such devices. While such a preparation is not perfectly suitable for performing fundamental investigations, it is relevant for understanding the properties of ionisation gauges.

## 2. Experimental procedures

### *2.1 Ionisation gauge simulator (IGS)*

The scheme of the ionisation gauge simulator is presented in Fig. 1. Its construction is very similar to Bayard-Alpert ionisation gauges, gas ionisers of residual gas analyzers (RGA) or ion sources based on electron impact ionisation. A circular hot filament, thoriated tungsten (ThW) or yttria coated iridium (YIr), is placed outside the positive cylindrical grid as an electron emitter. The filament is 5 cm long and 0.15 mm in diameter. The grid consists of 2 mm spaced 0.3 mm thick W vertical posts spot welded on a thin Mo disc and a Mo ring. Electrons are accelerated and enter into the grid volume, where they can ionize a gas introduced along the axial direction. The whole assembly is placed into the outer Mo cylinder, which is held on the filament potential. Ions are extracted axially from the grid volume through the aperture of bottom Mo disc by a stainless steel extraction electrode. The latter is a 7 mm thick disc with aperture diameter of 15 mm, placed 5 mm below the bottom of the grid. The insulating parts are made of aluminium oxide and macor.

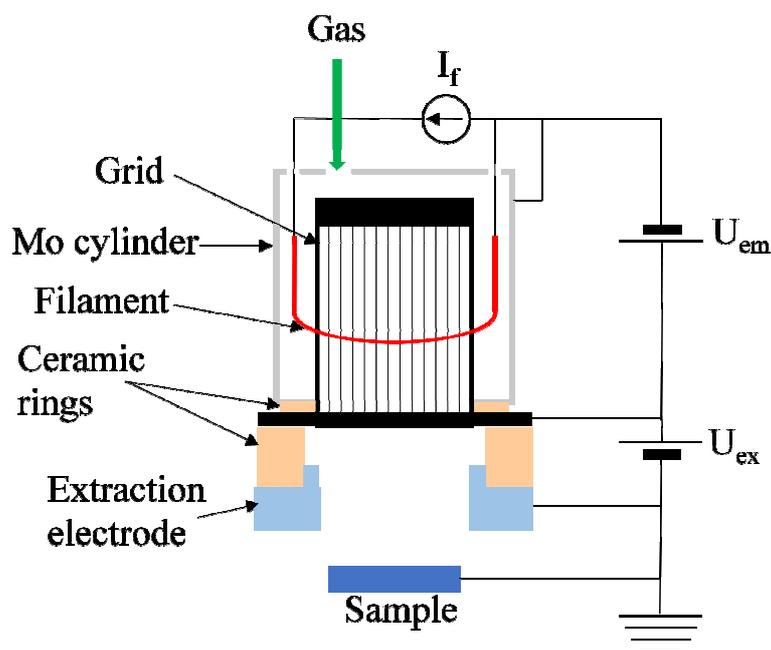

Figure 1. Schematic of the ionisation gauge simulator (IGS).

*2.2 Experimental setup*

IGS was assembled in the preparation chamber of an upgraded KRATOS XSAM 800 system produced by Kratos analytical [8], which also contains a quadrupole mass spectrometer for residual gas analysis and a sample holder allowing translation in three directions and rotation around the axis passing through the sample surface. The base pressure (i.e. with the IGS turned off) inside the preparation chamber was $2\cdot 10^{-8}$ mbar during these experiments.

Analysis chamber of this setup contains, among other devices, non-monochromatic X-ray gun with Al and Mg electrodes, electron energy spectrometer with a hemispherical energy analyzer (127 mm main path radius), an ion gun that can be used for sputter cleaning but also for IISEY measurements and sample annealing system by means of both thermal irradiation and electron bombardment. A sample transfer system allows introduction of samples and their transport between the fast entry lock chamber, preparation chamber and the analysis chamber without breaking the vacuum envelope. The base pressure in the analysis chamber is in the low $10^{-10}$ mbar range.

XPS measurements were performed using the Mg Kα X-ray source (photon energy of 1253.6 eV). Survey and high-resolution spectra were taken in a fixed analyzer transition mode with the pass energy of 40 eV and 20 eV, respectively. The binding energy axis was calibrated from the positions of the Au $4f_{7/2}$ line (83.96 eV) and Cu $2p_{3/2}$ line (932.4 eV) and the modified Auger parameter of copper (1851.2 eV) taken from sputter cleaned Au and Cu metallic samples. High resolution spectra were fitted using the pseudo-Voigt symmetric profiles after removing a background of Shirley type, unless otherwise stated. Surface composition was estimated using the atomic sensitivity factors from [9], corrected for the anisotropic part of the photoionisation cross section since the angle between the directions of the X-ray beam and the detected photoelectrons in this setup is 69°. This correction was made using the asymmetry parameters of photoelectron lines from [10]. Since quantification with sensitivity factors is applicable only for uniform samples, which was not the case here, the percentages obtained by this approach should be considered just to follow the general trends of surface modifications with sample treatments. In addition, we were applying in some cases another model suitable for in-depth non-uniform samples [11] to estimate thickness of a hydrocarbon overlayer at the processed samples.

Work function measurements were performed by monitoring low energy cutoff of the secondary electron emission spectra induced by X-ray irradiation of samples biased to -20.0 V. This cutoff is equal to the sample work function from which biasing voltage and spectrometer work function are subtracted [12]. While this approach cannot provide absolute value of the work function, one can monitor its change with the precision of few tens of meV [13].

Ion induced secondary electron yield measurements are performed in the analysis chamber of the Kratos setup, using a sputter ion gun. Ions are impinging the sample surface at 45° with respect to the surface normal. IISEY was calculated using the quotient method without a collector [14], i.e. from the currents measured on the biased sample irradiated by an ion beam, similarly to [15]. When the sample is biased to +15 eV, secondary electrons are recaptured so that the measured current $I^+$ equals the true ion current $I_p$. Negative biasing to -20 V repels the secondary electrons and prevents that tertiary electrons (electrons produced when secondary electron from the sample impinge the chamber walls) return to the sample. The corresponding current $I^-$ therefore equals the sum of $I_p$ and the secondary electron contribution $\gamma_{e,ion} \cdot I_p$. Consequently, IISEY can be calculated as

$$\gamma_{ion} = \frac{I_e}{I_p} = \frac{I^-}{I^+} - 1. \qquad (1)$$

Since the distance between the exit from the ion gun and the sample is about 20 cm, sample biasing has negligible effect on the ion current at the exit from the ion gun. Critical point in such a measurement is securing that ions primarily hit sample surface (and not e.g. the sample holder).

In that respect, particularly important part of the system upgrade is computer control of the ion beam deflection voltages, enabling formation of an image of a sample mounted on the sample holder: horizontal and vertical deflection voltages define the coordinates of a point in a graph, whilst the measured current determines its brightness. This functionality allows observation of the relative position of a beam spot at the sample surface, which allows mechanical positioning of a sample so that the beam hits its centre for zero deflection voltages. The latter condition secures that the position of the beam spot is not changing with the ion energy. Ion energies are estimated as the difference between the potential of the ion source anode and the negatively biased sample potential. All IISEY measurements were performed using $Ar^+$ ions in the energy range from 120 eV to 320 eV, where the lower part of this range is typical for ionisation gauges.

*2.3 Sample preparation and exposure conditions*

Several samples of potential interest in ionisation gauges were used in this investigation: polycrystalline molybdenum (Plansee, 0.15 mm thick foil, 10×10 $mm^2$, 99.97% purity) and gold (Goodfellow, 0.1 mm thick foil, 10×10 ×1 $mm^2$, 99.999%), as well as mechanically polished austenitic stainless steel (type 304, 0.5 mm thick, 10×15 $mm^2$). Besides, Cu(100) single crystal (MaTeck, 10 mm diameter, 1 mm thick) was also used for studying the mechanism of surface contamination in the ionisation gauge environment. After cleaning by isopropyl alcohol and distilled water in an ultrasonic bath, the copper sample was sputter cleaned before each exposure experiment in order to remove the surface impurities accumulated during a preceding treatment. Other samples were only cleaned in the same solvents as the copper sample. Such a procedure, not suitable for surface science experiments, was applied to the technical surfaces in order to achieve the initial conditions that could be expected at ion collector surface of an ionsation gauge. Due to the same reason, Mo and Au samples were not mechanically polished, aiming to preserve their original surface morphologies. In addition we were also studying graphite sample obtained by spraying a graphite lacquer GRAPHIT 33 (Kontakt Chemie) directly onto the stainless steel sample holder. The spray contains graphite powder and a lacquer dissolved in isopropanol. After drying in air, the sample was introduced in vacuum where it was annealed to about 140 °C for 30 min. This procedure removed efficiently the lacquer, as confirmed by XPS, leaving practically only carbon at the surface with very high $sp^2$ content.

Sample exposures to the IGS environment were performed by igniting a gas discharge at the pressure of $1.0 \cdot 10^{-5}$ mbar, measured by an ionisation gauge mounted in the preparation chamber. Most of the experiments were performed with Ar gas, although some exposures to $H_2$ gas discharge were also made. The gas line was pumped in the low $10^{-2}$ mbar range before introducing high purity gases (having a few ppm contamination level) to the pressure of about 1500 mbar. This procedure secured partial pressure of impurities in the low $10^{-10}$ mbar range, being about 2 orders of magnitude below the base pressure. The electron emission current was 5 mA, voltage between the grid and the filament was $U_{em}$ = 80 V and the grid potential was $U_{ex}$ = 200 V (cf. Fig. 1). Such conditions secure $Ar^+$ ion current density of 0.5 µA/cm$^2$ at the sample surface. In order to keep similar operating pressure as in the case of Ar discharge, the operating pressure, measured by a Bayard-Alpert gauge, was fixed to about $3.1·10^{-6}$ mbar due to ~2.8 times smaller gauge sensitivity for $H_2$ as compared to Ar. Ion current density of hydrogen ions achieved with these parameters was about 0.21 µA/cm$^2$. All exposures to IGS environment were performed under these conditions, unless otherwise stated.

In order to estimate at which pressure one could expect Ar ion current density of about 0.5 µA/cm$^2$, let us consider a hypothetical Bayard Alpert ionisation gauge operating with electron emission current $I_{em}$ = 1 mA having a sensitivity of S = 20 mbar$^{-1}$. If the ion collector has diameter of 0.1 mm and length of 10 cm, the total ion current would be about $I_i$ = 157 nA. Then, we estimate that this ion current would be achieved at a $p = I_i/I_{em}/S \approx 8·10^{-6}$ mbar.

Several sets of measurements, consisting of surface characterization (XPS, WF, IISEY), sample exposure and surface characterization, have been performed for all samples in order to check the reproducibility of the measurements. Occasionally, the samples were annealed and sputter cleaned by 1.5 keV $Ar^+$ ion beam in order to reduce the amount of surface hydrocarbons due to the previous exposures and/or air exposure.

## 3. Results and discussion

### 3.1 Gas composition in the operating IGS

After turning on the filament of the IGS the pressure rises for about one order of magnitude and then stabilizes to about $4·10^{-8}$ mbar. The residual gas mass spectrum with the IGS fitted with yttria coated iridium filament under steady state operational conditions is presented in Fig. 2. Mass spectrum taken right after the filament is turned off is also presented in the same figure. It should be noted that the mass spectra taken with tungsten filament are very similar both qualitatively and quantitatively.

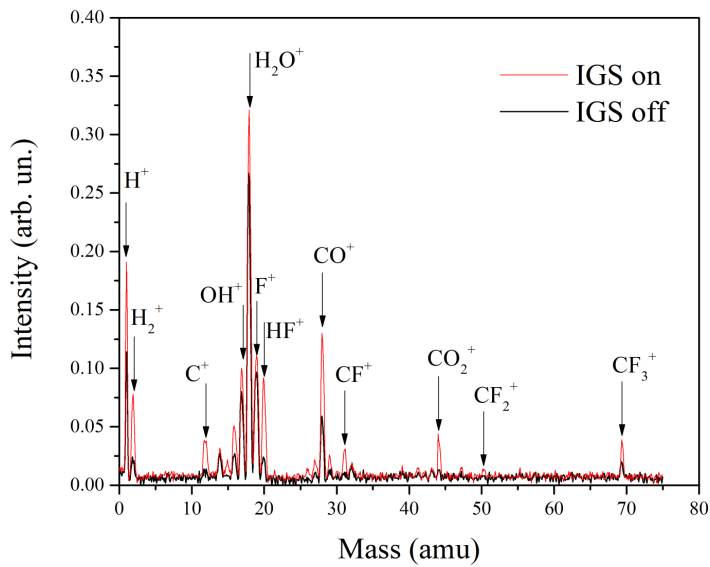

Figure 2. Residual gas mass spectrum in the preparation chamber with IGS filament turned on and off. The base pressure, measured by a Bayard-Alpert ionisation gauge was 3 to $4 \cdot 10^{-8}$ mbar.

Apart from the expected peaks in this pressure range attributed to water vapour, CO, $CO_2$ and hydrogen, we also see pronounced peaks at masses 19 ($F^+$), 31 ($CF^+$), 50 ($CF_2^+$) and 69 ($CF_3^+$), which we attributed to macor. The peak at 20 amu is most probably related to $HF^+$, although it can also be to some extent an artefact that is sometimes encountered in the case of intense $H_2O$ signal. Particularly interesting information can be obtained from the peaks with significantly reduced intensities right after the IGS is turned off. First of all, the intensities of $H^+$, $CO^+$ and particularly $H_2^+$ peaks strongly decreased. Peaks corresponding to masses 12 ($C^+$), 15 ($CH_3^+$), 29 ($C_2H_5^+$), and 44 ($CO_2^+$) are reduced to the noise level. Peaks related to the water vapour are also reduced, as well as masses 19, 20, 31, 50 and 69 that we attributed to fluorine species from macor. We therefore conclude that operation of IGS increases the partial pressure of hydrocarbon species, $H_2O$, CO and $CO_2$, which can be also expected in the case of any hot cathode gauges. In addition we observe some species characteristic to macor, hence specific to our experimental setup. Nevertheless, since the overall pressure rises by only $2 \cdot 10^{-8}$ mbar due to the IGS operation, relative contribution of these species to the overall pressure during the samples' exposure to Ar discharge is only 0.2 %, since the operating pressure was $1 \cdot 10^{-5}$ mbar during the sample exposures.

*3.2 Effect of exposure to IGS environment on surface composition*

In the first tests we performed XPS analysis of Mo, Au and stainless steel samples before and after an exposure to the IGS environment with Ar gas for 15 min. These experiments were performed with the ThW filament. The samples were not sputter cleaned prior to the experiment, which was suggested as conditioning procedure before a measurement with an ionisation gauge. XPS measurements reveal presence of carbon, oxygen and the corresponding metals at the surface. From the analysis of the O 1s lines and the characteristic photoelectron lines of the detected metals, we identified metallic oxides, pure metals and C-O bonds. The dominant contribution of the C 1s line was attributed to C-C and C-H bonds in saturated hydrocarbons accompanied with a minor contribution of C-O bonds shifted for about 1.5 eV towards higher binding energies [16]. The analyzed samples are typical examples of in-depth non-uniform surfaces, consisting of a metallic bulk, an oxide layer (except in the case of Au) covered by a layer of hydrocarbon impurities.

The results of the composition analysis are summarized in Table 1. High relative amount of carbon is common to all samples, implying relatively thick hydrocarbon overlayer on their top. After the IGS exposure, the dominant contribution of the C 1s line further increases, whilst the contribution of C-O bonds diminishes and eventually disappears after prolonged exposures to the IGS environments. This result implies that IGS exposure of all samples increases thickness of the hydrocarbon overlayer $d_c$, in which the amount of oxygen is becoming reduced with time.

Besides the increase of the hydrocarbon overlayer thickness, IGS exposure may also influence metallic oxide layer. We illustrate this on the example of the molybdenum sample. High resolution spectra of Mo 3d line taken before and after the IGS exposure are presented in Fig. 3. The main contribution in both spectra, with Mo $3d_{5/2}$ line at 228.3 eV, can be attributed to metallic molybdenum [17]. The secondary contributions correspond to $MoO_3$ (Mo $3d_{5/2}$ at 232.9 eV [17]) and $Mo_2O_5$ (Mo $3d_{5/2}$ at 231.7 eV [18]) in the spectra taken before and after the IGS exposure, respectively. Besides that, the relative contribution of the oxide phase in the Mo 3d line decreased from 25 to 21% by the exposure, implying that the oxide layer was partially sputtered away. This suggestion is further supported by the oxide phase reduction from Mo(VI) to Mo(V), which can be explained by well-known preferential sputtering of oxygen in different oxides of heavy metals (cf. [19] and references therein). Therefore, this result is a strong hint that additional hydrocarbon contamination during the IGS exposure is accompanied by $Ar^+$ ion sputtering of the sample surface.

Table 1. XPS composition analysis and position of the low energy cutoff of Au, Mo and stainless steel samples before and after the exposure to the IGS environment.

| Element | Gold | | Molybdenum | | Stainless steel | |
|---|---|---|---|---|---|---|
| | before | after | Before | after | before | after |
| C | 72.6% | 78.0% | 48.3% | 58.7% | 50.0% | 57.1% |
| O | 16.8% | 13.0% | 38.9% | 26.7% | 41.2% | 34.1% |
| Au | 10.5% | 8.9% | | | | |
| Mo | | | 12.8% | 14.6% | | |
| Fe | | | | | 5.2% | 5.3% |
| Cr | | | | | 3.5% | 3.5% |
| $E_{cutoff}$ (eV) | 19.71 | 19.97 | 20.12 | 20.01 | 20.34 | 20.00 |

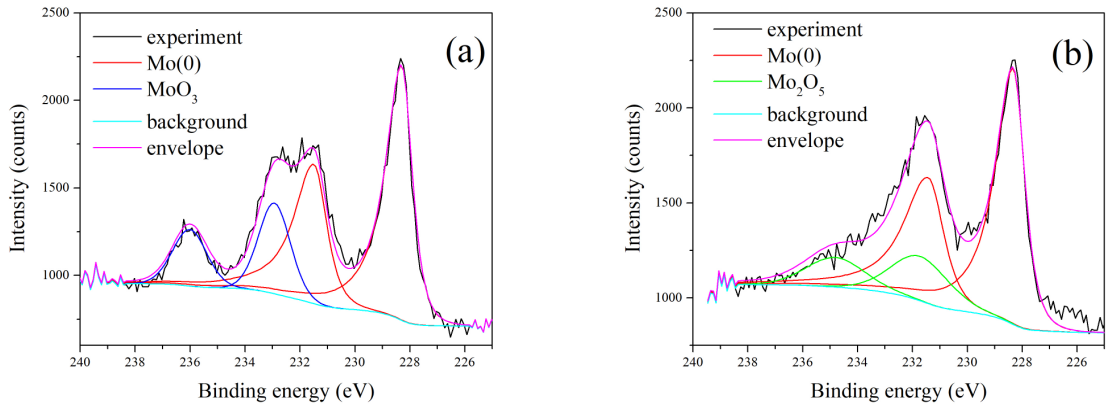

Figure 3. High resolution XPS spectra of Mo 3d line taken (a) before and (b) after the IGS exposure, and the corresponding line fittings.

In contrast to molybdenum, IGS exposure of stainless steel apparently increases the thickness of the oxide layer. We present in Fig. 4 fittings of the Fe $2p_{3/2}$ line taken from stainless steel sample before and after the IGS exposure. Since this line has very complex shape in the case of iron oxides due to the multiplet splitting effect, it is fitted as a superposition of symmetric pseudo-Voigt peaks with relative positions, intensities and widths constrained according to the measurements performed on reference compounds [20]. Therefore, although the line is fitted to the superposition of several peaks, the only free parameters per chemical phase are just the position and the intensity of the first peak. By applying this approach we identified two phases of iron: metallic iron, described by a single asymmetric peak (with a profile proposed in [20]) at 706.4 eV and FeO with the first peak at 709.3 eV (Fig. 4(a)) and 708.3 eV (Fig. 4(b)). The relative contribution of the metallic phase decreased from 24.4% to 12.4% after the IGS exposure, in contrast to the Mo sample. While the FeO contribution is at the expected position after the IGS exposure [20], this is not the case for the spectrum at Fig. 4 (a). Since the overall line shape fits much better to that of FeO than e.g. $Fe_2O_3$ and that we observe similar shift of the corresponding O 1s line contribution, we consider that the oxide layer is differentially charged due to the X-ray irradiation and limited conductivity of the oxide layer in this case.

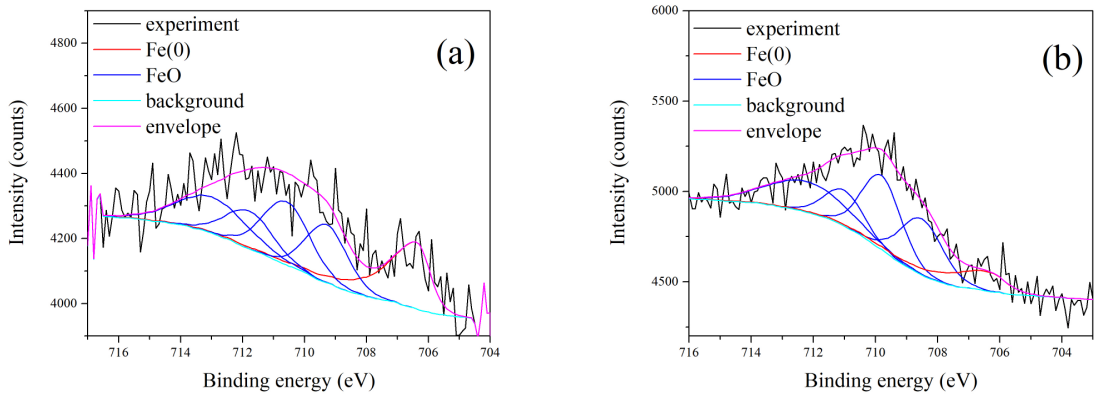

Figure 4. High resolution XPS spectra of Fe $2p_{3/2}$ line taken a) before and b) after the IGS exposure, and the corresponding line fittings.

Low energy cutoffs $E_{cutoff}$ of the secondary electron spectra measured before and after the IGS exposure are also presented in Table 1 for the exposed samples. The initial maximum difference between $E_{cutoff}$ of the three samples of 0.63 eV is strongly reduced after the exposure to only 0.04 eV, which is in the frame of the experimental uncertainty. The fact that the three work functions are becoming the same after the exposure indicates that the first layer of the three samples became very similar after the exposure. Having the same work function after the exposure regardless of the material would be particularly important if potential electron emission is dominant contribution to IISEY. The direct consequence would be that $\gamma_{e,ion}$ becomes independent of the electrode material. In addition, this result supports the XPS findings that all three samples are covered with a similar hydrocarbon layer after the exposure.

## 3.3 The mechanism of surface contamination

The results from the previous section show that exposures to environments like those in our experiments enhance surface contamination by saturated hydrocarbons, i.e. species that were also detected in the residual gas. At the same time we see on the example of Mo surface that ion sputtering of exposed surfaces, which might also contribute to hydrocarbon removal, cannot be ruled out. Therefore, a set of experiments was performed with the goal of revealing the mechanism of hydrocarbon contamination and particularly to understand the contribution of ion bombardment to the overall process. Copper single crystal was used as a sample due to its high purity and low surface roughness, which allows much more efficient sputter cleaning as compared with the samples used in previous experiments. The previously cleaned sample was exposed to a) Ar gas, b) Ar gas with the IGS filament on but without applying grid voltage that could accelerate electrons and provide gas ionisation and c) Ar gas with the IGS filament on and grid voltage applied (Ar discharge). The exposure time and the Ar pressure was the same in the three experiments, namely, 15 min and $1 \cdot 10^{-5}$ mbar, respectively. The aim of the experiment a) was to find the contribution of the gas residuals, from the preparation chamber and the gas line, to the overall surface contamination. In experiment b) the contribution of outgassing, mainly from the filament, was identified. In experiment c) the contribution of $Ar^+$ ion bombardment on the contamination process was expected to be revealed.

XPS measurements of the Cu sample were performed before and after the exposure. In each spectrum only lines characteristic to copper and carbon were observed, as well as some traces of fluorine. The sample surface can be modeled as a flat copper bulk covered by the thin film of saturated hydrocarbons. In order to provide proper quantification of such surface, the hydrocarbon layer thickness $d_c$ was determined rather than the surface composition using the atomic sensitivity factors. The former can be evaluated from the measured ratio of C 1s and Cu $2p_{3/2}$ line intensities $I_C$ and $I_{Cu}$ if the compositions of the overlayer and the bulk are known.

$$\frac{I_C}{I_{Cu}} = \frac{I_{C0}}{I_{Cu0}} \cdot \frac{1-e^{-\frac{d_c}{L_C^c}}}{e^{-\frac{d_c}{L_{Cu}^c}}}, \qquad (2)$$

where $I_{C0}$ and $I_{Cu0}$ are line intensities in the case of uniform hydrocarbon and copper samples, while $L_C^c$ and $L_{Cu}^c$ are effective attenuation lengths of C 1s and Cu $2p_{3/2}$ lines in hydrocarbon layer, respectively. $I_{C0}$ and $I_{Cu0}$ depend on the atomic concentration of carbon and copper, their photoionisation cross sections [21] and asymmetry parameters [10], as well as the transmission function of the energy spectrometer [8]. The parameters $L_C^c$ and $L_{Cu}^c$ were obtained from the corresponding NIST database [22] for the known composition of the contamination overlayer. Their values are 3.07 nm and 1.9 nm, respectively. Similar to [11], we model the layer of saturated hydrocarbons as paraffin, which allows us to determine all parameters in the transcendent eq. (2), thus leaving only $d_c$ as unknown magnitude. Therefore, by solving this equation, the change of $d_c$ due to the IGS exposure can be estimated from the measured ratios of carbon and copper line intensities.

Table 2. Change of the hydrocarbon layer thickness from the XPS results, calculated using the single layer model

| sample | Exposure | $t_{exp}$ (min) | $\Delta d_c$ (nm) |
|---|---|---|---|
| Cu | Ar gas at $1 \cdot 10^{-5}$ mbar | 15 | 0.05 |
| Cu | Ar gas at $1 \cdot 10^{-5}$ mbar with IGS ThW filament on | 15 | 0.2 |
| Cu | Ar gas discharge in IGS (ThW filament) at $1 \cdot 10^{-5}$ mbar | 15 | 1.83 |
| Cu | Ar gas discharge in IGS (ThW filament) at $1 \cdot 10^{-5}$ mbar | 60 | 4.58 |
| Au | Ar gas discharge in IGS (ThW filament) at $1 \cdot 10^{-5}$ mbar | 15 | 0.52 |
| Au | $H_2$ gas discharge in IGS (YIr filament) at $1 \cdot 10^{-5}$ mbar | 60 | 0.85 |

The results of the experiments are given in Table 2. Exposure just to the Ar gas with the ThW filament turned ON or OFF yielded in modest or negligible increase of $d_c$. However, acceleration of electrons, which enables ionisation and consequently ion bombardment of the sample surface, increases the rate of contamination for almost one order of magnitude! When the exposure time was four times longer, the estimated increase of $d_c$ was about 2.5 times greater. However, we should be aware that the method we apply to estimate $d_c$ has modest precision when the overlayer thickness becomes much greater than the effective attenuation lengths. The same effect was also observed in the case of gold, although the hydrocarbon growth rate was about 3.5 times slower. Effective attenuation length of Au $4f_{7/2}$ line in paraffin calculated using the NIST database was 3.6 nm. In addition, hydrocarbon layer was growing on gold also in the hydrogen discharge operating with YIr filament, although with even slower rate.

At first sight, enhancement of hydrocarbon layer growth by ion bombardment, instead of its suppression due to the sputtering, is a counterintuitive result. As the matter of fact, this is an effect known for a long time [23],[24],[25],[26],[27]. Surface adsorption of hydrocarbon species will inevitably take place when they are present in the residual gas, due to the high sticking coefficient and considerable desorption energy. When such surface is bombarded by a photon or electron beam, previously adsorbed hydrocarbon molecules will dissociate into free radicals. The latter will then make cross-links, thus forming chemically inert polymer-like structures. In the case of ion bombardment the two opposite effects will be taking place: removal of hydrocarbon species by ion sputtering and the enhancement of their polymerization. Which process will prevail depends on the sputtering yield and the partial pressure of hydrocarbons in the residual gas [23]. The flux of Ar ions impinging the surface, calculated from the ion current density, is about $3.1 \cdot 10^{12}$ $cm^{-2}s^{-1}$. Assuming the partial pressure of $CH_4$ of about $3 \cdot 10^{-9}$ mbar (about 10% of the total base pressure), the corresponding atomic flux will be $\sim 6 \cdot 10^{12}$ $cm^{-2}s^{-1}$ i.e. about 2 times higher than that of Ar ion flux. In addition, 200 eV Ar ions have too low energy to form collision cascades which is usually the dominant sputtering process. Momentum transfer to target atoms by direct knock-on collisions is still possible, but the incident angle is not suitable to promote this sputtering mechanism. Therefore, very low sputtering yield is expected, which all together results in enhancement of the hydrocarbon layer growth. The latter takes place even in the case of hydrogen discharge, although hydrogen atoms also perform chemical sputtering of hydrocarbons [28]. As a matter of fact, this is a possible reason why the hydrocarbon growth rate is lower in the case of hydrogen as compared to argon. In addition, it should be also noted that the current density of hydrogen ions was about 2.5 times smaller than that of Ar ions (cf. Section 2.3).

It should be emphasized that the residual gas composition can have significant influence on the hydrocarbon layer growth. Usual sources of carbon in vacuum systems are hot tungsten filaments [29]. Carbon atoms, dissolved in the bulk of the filament, segregate to the surface at high temperature. These atoms then react with the residual gas, thus forming volatile species such as $CO$, $CO_2$ and small hydrocarbon molecules. Formation of hydrocarbon in vacuum can be closely related with the original residual gas composition: presence of hydrogen precursors, such as $H_2$ or $H_2O$, will promote formation of hydrocarbons. Therefore, in systems with smaller amounts of hydrogen precursors, the hydrocarbon growth rate can be reduced and even suppressed. If, on the other hand, there are other hydrocarbon species in the residual gas (e.g. different from small alkane molecules), surface exposure to other hydrogen precursors (namely $H_2$ and $H_2O$, which are clearly present in the residual gas) might also contribute to the formation of saturated hydrocarbons at the sample surface.

When a sample is not homogeneous in depth, which we have in the case of hydrocarbon layer covering all samples including gold and copper, other effects may play a role. It is well known that sputtering yield is directly proportional to the ion energy deposited practically in the first nanometer. If atomic mass of bulk atoms is much larger than that of argon, as in the case of Au, ion energy is efficiently deposited into the hydrocarbon layer. Consequently, the sputtering yield of hydrocarbons will be enhanced, which is known as sputtering yield amplification [30].

These conditions are not met in copper sample: Ar ions will efficiently transfer energy to Cu atoms due to the more similar masses, thus reducing energy confinement into the hydrocarbon overlayer. Therefore, sputtering yield amplification could be responsible for the smaller growth rate (i.e. higher sputtering rate) of the hydrocarbon layer on Au as compared to Cu.

*3.4 The consequences of the IGS exposure on IISEY*

Knowing that secondary electron emission induced by low energy ions is related to first few atomic layers of the surface, it is to be expected that IGS exposure will change IISEY. Time evolution of $\gamma_{e,ion}$ may be followed particularly well on surfaces that can be efficiently cleaned e.g. by vacuum annealing, such as graphite, which we consequently tested. Prior to the exposure experiments, the graphite sample was annealed to 350 °C for 20 min, yielding in a high purity surface containing about 1% of O. In Figs. 5 (a) and (b) we present high resolution spectra of the C 1s photoelectron line taken before and after the IGS exposure to Ar discharge ($p_{Ar} = 1·10^{-5}$ mbar) for 15 min.

The peak fitting model was built by fitting the C 1s line taken from the highly oriented pyrolytic graphite, which was used as a reference for pure graphite: we apply Doniach-Šunjić profile with the asymmetry parameter of 0.01, convoluted by pseudo-Voigt profile in the form of product of Gaussian (30%) and Lorentzian (70%) (having a standard notation GL(70)). C 1s line fitting confirms very high purity of the freshly annealed sample. About 91% of carbon atoms make $sp^2$ i.e. pure graphitic bonds, accompanied with 5% of $sp^3$ and 4% of C-O bonds. Finally, a broad peak at about 290 eV, representing $\pi$-$\pi^*$ shake up satellite, is another fingerprint of the graphitic structure.

After the IGS exposure the relative amount of carbon decreases to 88%, accompanied with equal relative amounts of oxygen and fluorine. The latter is present in the form of C-F bonds (F 1s line is situated at 688.7 eV) [31]. However, the greatest change was observed in C 1s line (Fig. 5(b)). The relative amount of $sp^2$ dropped to only 54%, whilst a contribution that we attribute to C-C and C-H bonds in saturated hydrocarbons increased to 32.4%. C-O contribution at 286.3 eV [31] increased to 8%, whilst additional contributions appeared at 287.8 eV (3.5%) and 291 eV (2.2%). They can be attributed to C=O and $CF_3$ bonds, respectively [31].

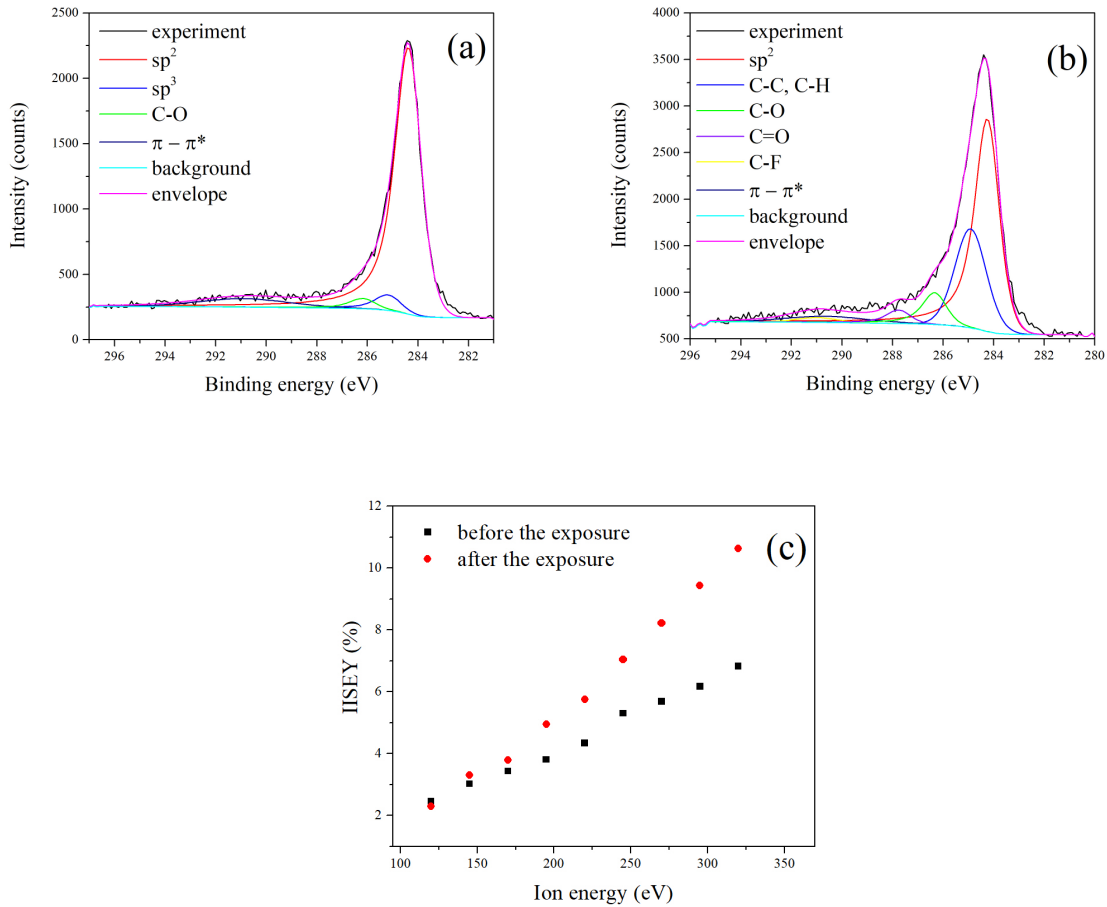

Figure 5. High resolution XPS spectra of the C 1s line taken from the graphite sample a) before and b) after the IGS exposure; c) energy dependence of IISEY of $Ar^+$ ions from the graphite sample before and after the IGS exposure.

Highly reproducible IISEY data for $Ar^+$ ions taken from the graphite sample before and after the IGS exposure are presented in Fig. 5(c). First of all, we notice obvious dependence of $\gamma_{e,ion}$ on the ion energy, which proves that the dominant process is subthreshold kinetic electron emission. The two sets of IISEY points practically overlap up to 150 eV. At higher energies, $\gamma_{e,ion}$ from the exposed sample becomes distinctly higher. This energy could represent a threshold for an additional channel for secondary electron emission. Existence of an energy threshold is another fingerprint of kinetic electron emission: it is related to the distance of closest approach between two collision partners at which a particular electron promotion process becomes possible.

Measurement of $Ar^+$ induced secondary electron yield from clean highly oriented pyrolytic graphite at Utah state university show very different result: in this energy region $\gamma_{e,ion}$ is practically energy independent, being about 25±15%, evidencing dominance of potential electron emission [32]. Significant discrepancy of the two measurements can hardly be explained by different measurement geometries: as previously stated, incident angle of the $Ar^+$ beam in our measurements was 45°, in contrast to normal incidence in [32].

IISEY measurements induced by $Ar^+$ ions from Au and Mo samples contaminated by IGS exposure and subsequently annealed (cf. next section) are presented in Figure 6. As in the case of the graphite sample, energy dependence of $\gamma_{e,ion}$ is clearly pronounced, evidencing strong contribution of the kinetic electron emission. $Ar^+$ induced secondary electron emission from clean Mo surface in this energy region is practically energy independent, typical for potential electron emission, having $\gamma_{e,ion}$ of about 11% [33]. Other groups report slightly lower values (~8% [34] and ~9% [35]). The same trend was observed in the case of $\gamma_{e,ion}$ of $Ar^+$ from gold, although IISEY is only about 5% in the same energy range [4]. In the case of tungsten, another metal frequently used for ion collectors, low energy $\gamma_{e,ion}$ equals about 9% and shows the same energy independence [33].

Review of various experimental results reveals that hydrocarbon layer growth on clean metals contributes to energy dependent IISEY: for energies typically below ~200 eV $\gamma_{e,ion}$ is lower, while at higher energies the yield is higher than that of clean metals [36]. The latter is consistent with our results. It should be also noted that $\gamma_{e,ion}$ of Mo at 220 eV in Fig. 6 is 11.3±0.5%, being close to the low energy values obtained for the clean metal. Finally, increased slope of $\gamma_{e,ion}(E)$ from graphite after the surface contamination (i.e. due to the sample exposure to the Ar discharge), as can be seen in Fig. 5c, further supports our assumption that pronounced kinetic electron emission is caused by the hydrocarbon layer growth. While the slope of $\gamma_{e,ion}(E)$ is approximately the same for contaminated Mo, Au and graphite surfaces, all the values of graphite are lower by about 4-5%. A possible explanation could be dissimilar contributions of the potential electron emission.

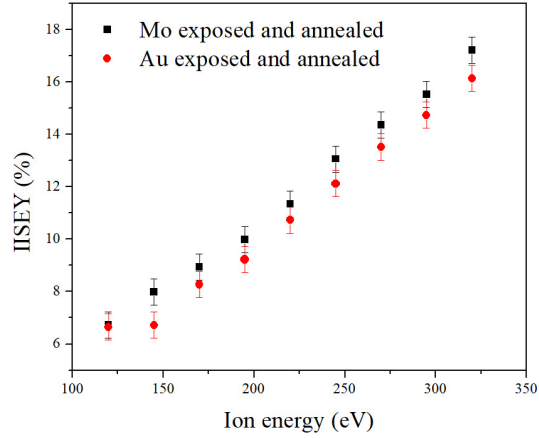

Figure 6. Energy dependence of IISEY of $Ar^+$ ions from Mo and Au samples previously contaminated by IGS exposures.

According to the XPS analyses, the Au and Mo samples can be approximated by a metallic bulk covered with a hydrocarbon layer. By applying the single layer model introduced in Section 3.3, we estimate that $d_c$ on Au and Mo samples is about 3.9 and 2.8 nm, respectively. IISEY data practically overlap for the two samples in the frame of the experimental error (although the points of gold are systematically below those of molybdenum), which is consistent with the surface nature of the electron emission and our previous finding that different sample surfaces are becoming similar after the IGS exposure. The only difference is the hydrocarbon growth rate, which appears to be lower for more inert surfaces (e.g. gold or graphite).

Significance of the observed $\gamma_{e,ion}$ change of materials by their exposure to Ar ion bombardment is in the fact that namely this kind of treatment is frequently used to improve the stability of the pressure reading in ionisation gauges [37]. It is generally assumed that this empirically established procedure contributes to surface cleaning by intense sputtering of surface impurities from ion collectors. Unexpectedly, its role appears to be in promoting the growth of a stable hydrocarbon layer. Having in mind that the inelastic mean free path of electrons with energy below 10 eV is rather large, constant $\gamma_{e,ion}$ will be reached when the layer is sufficiently thick to secure that all secondary electrons are emitted exclusively from this region. Since the ion range increases with the ion energy and decreases with the atomic number, the thickness should be larger than the range of the most energetic $H^+$ ions. When this thickness is achieved, standard operation will contribute to further growth of the hydrocarbon layer, but leave the ion induced secondary electron yield constant. This phenomenon will be taking place in any ionisation gauge, regardless of its design or the electrode materials. Related to our experiments, we have performed Monte Carlo simulations of 100 and 300 eV Ar ions impinging paraffin (32.5% C, 67.5% of H, density of 0.96 g/cm$^3$) using well known SRIM code [38], version SRIM-2013. It appears that all Ar ions stopped within first 3 nm (implantation profile maximum at 1.4 nm) and 5 nm (implantation profile maximum at 2.3 nm) in the case of 100 eV and 300 eV ions, respectively. Small difference between the evaluated Ar ion ranges in paraffin and the estimated thickness $d_c$ of Au and Mo samples from which IISEY measurements were performed explain rather small difference between the electron yields.

*3.5 Influence of sample annealing on the hydrocarbon layer*

Another issue of interest in vacuum technology is how vacuum baking affects the hydrocarbon layer previously grown on the surface of an ion collector. In that respect we were determining change of $d_c$ on Au and Mo samples due to the sample annealing from the XPS analyses and the previously introduced single layer model. Each annealing was performed at 350 °C for 20 min. Assuming that organic species are non-dissociatively desorbed according to the Langmuir model with the desorption energy of 1.5 eV, their desorption rate at this temperature is four orders of magnitude greater than at 180 °C (the latter being a typical vacuum baking temperature). Thus, we can roughly say that annealing a sample for 20 min at 350 °C is equivalent to vacuum baking for 20×10$^4$ min (≈139 days) at 180°C. Both samples were previously exposed several times to the Ar discharge, yielding to initial hydrocarbon layer thicknesses of about 6 nm (Au) and 4.5 nm (Mo). This annealing procedure decreased $d_c$ of Au and Mo to about 4 nm and 3.3 nm, respectively. Repeated annealing of Mo under the same condition reduced further $d_c$ to 2.8 nm, whilst additional annealing cycles did not show any effect on the hydrocarbon layer. At the same time, IISEY measurements performed between the annealing cycles were identical in the frame of the experimental error.

From these experiments we can conclude that hydrocarbon layers formed at metallic surfaces are too stable to be affected by standard baking procedures. The latter is well known in experimental surface science: the only efficient way to remove hydrocarbons from a surface is by ion sputtering. Consequently, once an ion collector is exposed to the operating conditions in an ionisation gauge, its $\gamma_{e,ion}$ should not be strongly affected by vacuum baking and/or low energy ion bombardment.

It should be stated, however, that the latter is not true for highly inert surfaces such as graphite. According to our experience, graphite sample previously exposed to IGS environment is efficiently cleaned by annealing at 350 °C for 20 minutes.

## 4. Conclusion

Our investigation showed that, depending on the material, its surface conditions and the ion energy, $\gamma_{e,ion}$ of an ion collector induced by $Ar^+$ ions can have values in the range 3-20%. The first direct consequence of this fact is that experimental determination of ionisation gauge sensitivity may be strongly affected by secondary electron emission since the true current of collected ions cannot be measured in typical ionisation gauge arrangements without a suppressor grid in front of a suitably designed ion collector [39]. Thus, comparison of sensitivities obtained theoretically or by simulations with experimental ones has to take into account this unknown quantity.

The most important conclusion of this research is that exposure of electrodes to an environment characteristic to ionisation gauges at room temperature inevitably leads to the growth of hydrocarbon layers. The contamination is triggered by particle bombardment of electrode surfaces: electrons and ions in the case of the grid and the ion collector, respectively. Such process is clearly observable as a brown colour of electrodes in ionisation gauges operating for only few tens of hours. Formation of a few nanometer thin hydrocarbon layer on the surface of an ion collector will already change $\gamma_{e,ion}$ sufficiently to affect the gauge sensitivity. This can potentially produce a drift in the reading of ionisation gauges during the process of the hydrocarbon layer growth. Growth of hydrocarbons seems to take place regardless of the discharge gas. It was experimentally observed in highly inert (Ar) or hydrogen rich atmosphere, although exposure to the latter should contribute to chemical sputtering of hydrocarbons.

Vacuum baking up to 200 °C or $Ar^+$ bombardment will not remove the hydrocarbon layer formed at the ion collector surface. On the contrary, the latter will probably further increase its growth rate. Another procedure of cleaning ionisation gauge electrodes, provided by many ionisation gauge controllers, is "degassing". In the case of the ion collector, it represents electrode heating by electron bombardment, reaching very high temperatures that will likely affect the hydrocarbon layer formed at the surface of an ion collector. The consequences of degassing on IISEY of ion collectors will be studied in a later work.

In spite of different cleaning procedures of ionisation gauges, hydrocarbon layers will be re-established during the normal operation of a gauge. This work shows that accuracy of ionisation gauges does not depend on how 'clean' is the ion collector surface, but how stable is its IISEY. Assuming that the surface morphology of an ion collector is not changing during the operation, its $\gamma_{e,ion}$ will likely be constant and independent of the electrode material and the general design of the ionisation gauge once a uniform and sufficiently thick hydrocarbon layer is formed. Possible consequence of the formation of such layer could be a more stable sensitivity, since we found that sufficiently thick hydrocarbon layers have relatively stable secondary electron yield. Since the growth rate of hydrocarbon species seems to be increasing with the surface reactivity, ion collectors made of transition metals should have shorter transition period of unstable secondary electron yield as compared to inert materials such as gold or graphite. In addition, it can be recommended that before measurement with an ionisation gauge, after a bake-out a conditioning procedure of operating the gauge for at least an hour at a few mPa Ar is introduced. This procedure builds up a hydrocarbon layer and stabilizes the ion induced secondary electron yield and the reproducibility of the sensitivity of the gauge.

Alternatively, stable $\gamma_{e,ion}$ could be achieved by using chemically inert electrodes operating at high temperature. Both, high temperature and low reactivity of the substrate will reduce equilibrium surface concentration of adsorbed hydrocarbon species. Consequently, the probability of formation of free radicals by ion bombardment, which can then cross link and polymerize, will be much lower. In other words, the conditions for the hydrocarbon layer growth will not be fulfilled under such circumstances, contributing to stable $\gamma_{e,ion}$ during the gauge operation (equal to that of a clean surface). This strategy was indeed applied in [40], where high stability of an ionisation gauge with gold coated electrodes operating at 250 °C was reported.

## Acknowledgments

This work has received funding from the EMPIR programme (project 16NRM05 'Ion gauge') co-financed by the Participating States and from the European Union's Horizon 2020 research and innovation programme, and the Portuguese National Funding Agency for Science, Research and Technology in the framework of the project UID/FIS/00068/2019. The authors also thank Mihail Granovskij and other colleagues from VACOM, for providing the yttria coated iridium filament used in our experiments.

**Conflicts of interest:** "The authors declare no conflict of interest."